\documentclass[runningheads]{llncs}

\usepackage{bibnames}
\usepackage[T1]{fontenc}
\usepackage[american]{babel}
\usepackage{graphicx}
\usepackage{lscape}
\usepackage{rotating}
\usepackage{csquotes}
\usepackage{amsmath}
\usepackage{amsfonts}
\usepackage{todonotes}
\usepackage[capitalise]{cleveref}

\def\eg{e.g., }
\def\metaspi{$\mu SPI$}

\begin{document}
%
\title{Towards an Argument Pattern for the Use of Safety Performance Indicators}
%
%
\author{Daniel Ratiu\inst{1} \and
Tihomir Rohlinger\inst{1,3} \and
Torben Stolte\inst{2} \and
Stefan Wagner\inst{3,4}
\institute{CARIAD, Munich, Germany \and
Volkswagen ADMT, Wolfsburg, Germany \and
University of Stuttgart, Stuttgart, Germany
\and
Technical University of Munich, Heilbronn, Germany}
}
\authorrunning{D. Ratiu, T. Rohlinger, T. Stolte, S. Wagner}
\maketitle              
\begin{abstract}
Highly automated driving functions pose challenges for the safety assurance due to their high complexity and the dynamic environment in which they operate.
UL 4600, the safety standard for autonomous products, mandates the use of Safety Performance Indicators (SPIs) to continuously ensure the validity of safety cases by monitoring and taking action when violations are identified. Despite numerous examples of concrete SPIs available in the standard and companion literature, their contribution rationale for achieving safety is often left implicit.
In this paper, we present our initial work towards an argument pattern for the use of SPIs to ensure validity of safety cases throughout the entire lifecycle of the system. Our aim is to make the implicit argument behind using SPIs explicit, and based on this, to analyze the situations that can undermine confidence in the chosen set of SPIs. 
To maintain the confidence in SPIs' effectiveness, we propose an approach to continuously monitor their expected performance by using meta-SPIs. 

\end{abstract}

\keywords{Safety Performance Indicators \and UL-4600 \and Safety Argument}

\section{Introduction}
Software-defined vehicles are fully connected and have complex software that takes over critical vehicle functions such as highly automated driving. 
This poses both opportunities and challenges to the traditional safety approaches. 
On the opportunities side, connectivity allows real-time and large-scale supervision of systems' performance and timely detection of safety-related insufficiencies. 
Once insufficiencies are detected, proactive measures can be implemented and deployed to the fleet over the air in the form of software updates before losses occur. 
On the challenges side, the high complexity of the software, short and iterative development cycles with continuous updates, changes in the environment unforeseen at design time, unforeseen edge cases and surprises, are all causes that can lead to invalidation of the safety case.
To handle these situations, there is a broad agreement in the safety community that the validity of the safety case needs to be continuously monitored
\cite{2020_fayollas_safeops_continuous_safety,2021_gyllenhammar_ads_safety_assurance_future_directions,2023_hawkins_identifying_runtime_monitoring_Requirements_for_autonomous_systems_through_analysis_of_safety_arguments,2022_johannsson_continuous_learning_approach_to_safety_engineering,2024_wagner_open_autonomy_safety_case_framework}. 

The UL~4600 standard \cite{2023_ul4600} for ensuring safety of autonomous products mandates the use of \emph{Safety Performance Indicators} (SPI) as a means to continuously validate the safety case based on the data collected from the fleet and perform counter-measures when violations are detected. 
SPIs collect counter-evidence empirically during the entire lifecycle of the product with a focus on operations. 
Despite the promising approach to continuously ensure safety using SPIs, its limitations, and challenges to deploy on a large scale are not well studied. 
There are different questions that need to be answered when the SPIs are used to ensure the validity of a safety case as exemplified below:

\begin{itemize}
    \item Does the chosen set of SPIs sufficiently cover the safety case?
    \item Is the handling of SPI violations sufficient to prevent losses?
    \item How can we validate that a certain set of SPIs adequately prevents the invalidation of the augmented safety case?     
\end{itemize}

\begin{figure}[ht]
\centering
\includegraphics[scale=0.5]{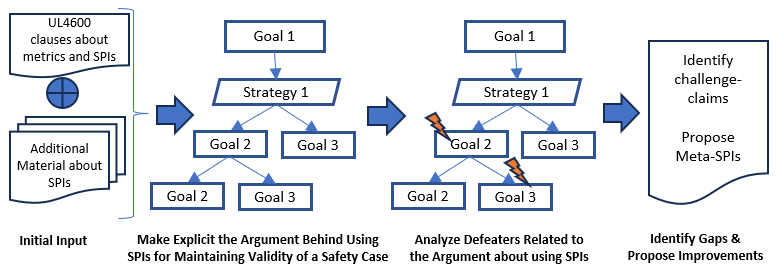}
\caption{Our approach at a glance.}
\label{fig:approach_overview}
\end{figure}

\paragraph{Approach.} In \cref{fig:approach_overview}, we present an overview of our approach to tackle these questions. 
Our starting point is an analysis of the UL~4600 \cite{2023_ul4600} standard itself regarding the use of metrics and SPIs and additional published material which explains the standard \cite{2023_koopman_ul4600_what_to_include_in_an_autonomous_vehicle_safety_case,2022_koopman_how_safe_is_safe_enough}.
We develop a top-level argument that makes explicit the claims made when using SPIs to maintain the validity of safety cases.
The argument about the use of SPIs is \enquote{just} another argument, and thereby, its weakness needs to be analyzed, and the confidence in its validity requires continuous validation.
After developing the argument, we analyze its claims for potential defeaters by using dialectical argumentation \cite{2023_hawkins_identifying_runtime_monitoring_Requirements_for_autonomous_systems_through_analysis_of_safety_arguments} and make explicit potential gaps using challenge claims. Finally, we propose the use of \emph{meta-SPIs} to validate a concrete SPI framework used to augment a safety case.

\paragraph{Structure.}
In \cref{sec:background}, we present background concepts related to assurance cases and UL~4600, which will be used in the next sections of the paper. 
\cref{sec:implicit_safety_case} presents a high-level argument structure about using SPIs for maintaining the validity of a safety case. 
\cref{sec:critical_analysis} contains a discussion about weak points and challenges of the argument behind using SPIs. In Section~\ref{sec:meta_spis} we propose the use of meta-SPIs for continuous validation of a concrete set of SPIs. 
\cref{sec:related_work} contains the related work and \cref{sec:conclusions} concludes the paper.


\section{Background}\label{sec:background}

\paragraph{Structured Safety Arguments,}  \eg modeled by using the \emph{Goal Structuring Notation} (GSN)~\cite{2021_scsc_gsn_community_standard_v3}, are used to capture the safety case as a hierarchy of claims, sub-claims, and evidence in the form of a tree. 
The root node of the tree is called \emph{top-level goal} and goals closer to the root are called \emph{high-level goals} whereas goals closer to the bottom of the tree are called \emph{low-level goals}. 
%
The validity of \emph{lowest-level goals} is demonstrated by evidence whereas the validity of the other goals is argued by their decomposition into sub-goals according to an explicitly stated strategy. A sub-argument in the argument tree is called \emph{argument leg} or \emph{argument branch}. A goal $G_i$ has a higher level than a goal $G_j$ if it is on the path in the tree between $G_j$ and the top-level claim. 

Violations of the top goal are associated with mishaps and losses and are thereby of the highest impact since they invalidate the safety case as a whole. Therefore, during the entire lifecycle of a safety critical system, it is necessary to prevent the violations of top claims by all means and thereby keep the system safe. Violations of low-level goals might happen more often and are not as severe as violations of high-level claims due to the different safety mechanisms and assurance margins which prevent sub-system-level deficits from propagating to system-level losses and mishaps.

\emph{Assurance deficits} of a safety case are gaps in the argument or evidence that potentially lead to the invalidation of the \emph{top claim}. Assurance deficits can occur due to weak evidence associated with leaf goals, the incomplete decomposition of goals into sub-goals, or violated design time assumptions at runtime, for example, when the environment changes.


\paragraph{UL4600 \cite{2023_ul4600}} is a safety standard for autonomous products. The core of the standard is formed by chapters 5--17, which provide the content of the safety case. The structure of these core chapters is formed by a list of numbered clauses, supported by additional normative phrases (called \enquote{prompt elements}), which provide more details about what should be addressed by the safety case. The prompt elements are classified as mandatory, required, highly recommended, recommended, or pitfall.

\emph{Safety Performance Indicators (SPIs)} are mandated by the UL~4600 standard as a means to link safety cases with runtime information to detect potentially ineffective risk mitigation planned and implemented at design time. UL~4600 uses the following definition of SPIs \enquote{\emph{An SPI is a metric supported by evidence that uses a threshold comparison to condition a claim in a safety case.}}\cite{2023_ul4600}. 
The metric values are empirically obtained through the observation of the product during its lifecycle. These values are compared to a threshold, and thereby, a predicate is obtained, which is true when the values are above or below the threshold. Each SPI is linked to a claim in the safety case and the violation of the SPI's threshold (i.e., SPI is false) means that the claim is violated. However, the claim can be violated even without the associated SPIs being false.

The safety performance indicators are categorized into lagging and leading indicators~--~\emph{lagging indicators} are used to detect shortcomings in the safety case based on the events that happened in the past (\eg system level incidents), whereas \emph{leading indicators} are used to identify potential safety issues (\eg unexpected activations of fail-safes) before severe loss events occur and thereby the leading indicators have a prediction power.
UL~4600 (note 16.2.2.6.2) explicitly states that relying solely on lagging indicators is not acceptable, and proactive indicators, which predict possible mishaps before happening, need to be employed.  
%
The use of SPIs is key to the UL~4600 standard, and they are mentioned throughout the entire standard. Below is an overview of the sections of the standard directly referring to SPIs and their metrics.

\begin{itemize}
    \item Section 5 (\enquote{Safety Case and Arguments}), in the context of clause 5.4.3 about the support of evidence validity for difficult-to-reproduce aspects of the item, the prompt 5.4.3.2 requires the use of SPIs; in context of the clause 5.6.1 about the role of safety culture in risk mitigation, the prompt 5.6.1.3 highly recommends the use of metrics for evaluating the safety culture.
    \item Section 6 (\enquote{Risk Assessment}), in clause 6.5.1 about risk mitigation methods, prompt 6.5.1.3 highly recommends the use of leading metrics to estimate post-mitigation risks and a comparison between leading metrics predictions and lagging metrics results to continually re-calibrate the leading metrics.
    \item Section 9 (\enquote{Software and Systems Engineering Processes}), clause 9.1.4 regarding the incorporation in the development process of best practices for safety-related elements, the note 9.1.4.6.3 mandates that configuration management needs to associate the SPI data to its corresponding configuration item for which it is collected. 
    \item Section 14 (\enquote{Lifecycle Concerns}), clause 14.6.2 mandates the mitigation of risks related to software updates, and prompt 14.6.2.2 requires the consideration of the updating of SPIs as part of the update safety impact analysis.
    \item Section 16 (\enquote{Metrics and Safety Performance Indicators (SPIs)}), is completely dedicated to the definition of metrics and SPIs and contains eight clauses. It starts with clause 16.1.1, which mandates the incorporation of SPIs in the safety case, and continues with five clauses, 16.2.1-16.2.5, about the definition of the safety performance indicators. Clause 16.3.1 is about the collection of SPI violations and metric data, and clause 16.3.2 is about the item improvement based on the SPI data. Clause 16.3.3 is about using non-SPI data to validate and improve the prediction power of SPIs. 
\end{itemize}


\emph{Use-Cases for SPIs.} SPIs are a generic instrument for monitoring the system and collecting outliers relevant for the safety case. Once this framework is in place and data is available, it can be used for different purposes such as 1) ensure continuous validity of the safety case throughout the entire lifecycle of the system,
2) making informed decisions about when to deploy, 
3) achieving and sustaining societal trust in the autonomy stack by transparency about the safety performance. In this paper, we focus on the first use case and provide a high-level argument structure for using SPIs to ensure continuous validity of the safety case, analyze its core challenges, and sketch a possible way forward.

\section{A High-level Argument Structure for Using SPIs}\label{sec:implicit_safety_case}

In this section, we present a high-level argument structure that makes explicit the implicit argument for using SPIs to maintain the validity of safety cases. Our longer term aim is to evolve this high level argument into a pattern. However, due to space limitations,  we use a simplified notation for presenting our argument instead of the standardized GSN extension for patterns. 
In \cref{sec:critical_analysis}, we analyze this argument with respect to potential defeaters, and in \cref{sec:meta_spis}, we propose a set of (meta-)SPIs, which validate arguments containing SPIs.

\begin{figure}[h]
\centering
\includegraphics[scale=0.5]{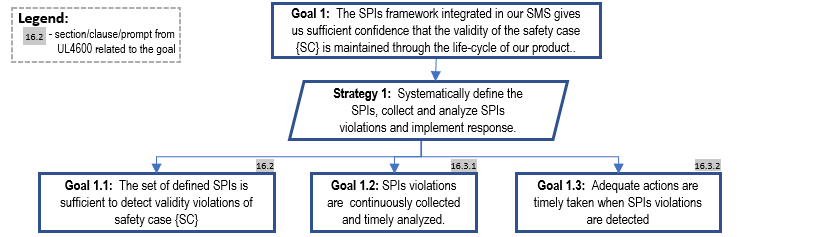}
\caption{Top-level argument about the use of the SPIs framework for maintaining the validity of a safety case \{SC\} through the life-cycle of a product.}
\label{fig:implicit_spis_safety_case_top_level}
\end{figure}

\paragraph{Top Level.} The top-level part of our argument is presented in \cref{fig:implicit_spis_safety_case_top_level}. The top claim ({\tt Goal 1}) in this argument states that the SPIs framework is integrated in the Safety Management System (SMS) and is used to maintain the validity of a given safety case {\tt SC} throughout the entire lifecycle of the product. {\tt Goal 1} is decomposed with the help of a strategy for the systematic definition, collection, analysis, and response to SPI violations as mandated in the UL~4600 standard. 
Three sub-goals ({\tt Goals 1.1--1.3}) develop each of these aspects, and their supporting argumentation legs are presented in \crefrange{sec:definition_of_spis}{sec:response_to_spis_violations}. In the following GSN diagrams we trace the goals related to UL4600 sections/clauses/prompts by using a shaded number on the top-right of the goal's rectangle -- \eg the {\tt Goal 1.1} is traced to section 16.2 from UL4600\cite{2023_ul4600}.

\subsection{Systematic Definition of SPIs} 
\label{sec:definition_of_spis}
\begin{figure}[h]
\centering
\includegraphics[scale=0.53]{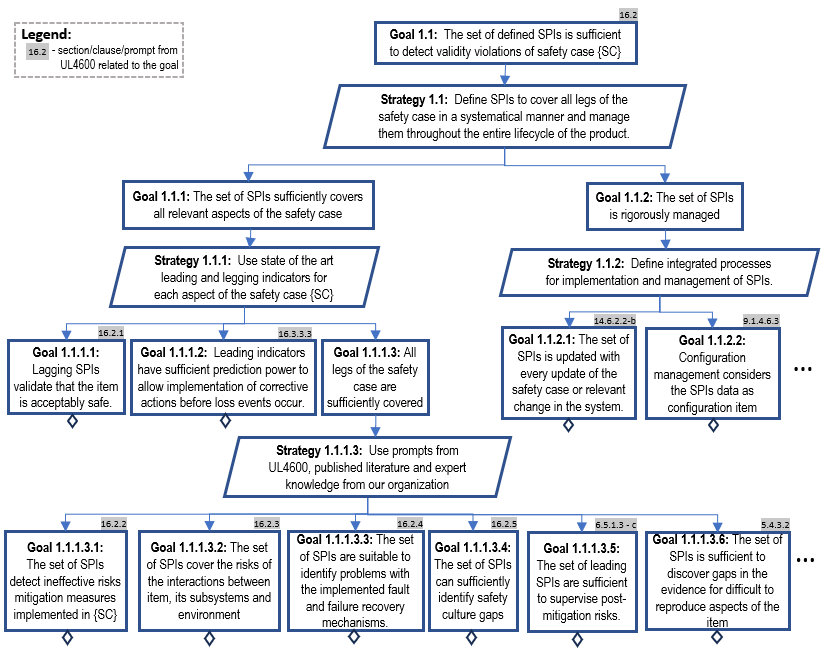}
\vspace{-0.4cm}
\caption{Argument leg about the definition of SPIs.}
\vspace{-0.6cm}
\label{fig:implicit_spis_safety_case_definition_of_spis}
\end{figure}

Having a safety case with claims annotated with SPIs, safety engineers need to ensure that the set of SPIs is sufficient to address its potential assurance deficits ({\tt Goal 1.1}). The argument leg about the sufficiency of SPIs definition, presented in \cref{fig:implicit_spis_safety_case_definition_of_spis}, has two aspects covered each by a sub-goal: adequate coverage of the safety case ({\tt Goal 1.1.1}) and the rigorous implementation and management of SPIs ({\tt Goal 1.1.2}).

\vspace{-0.3cm}

\paragraph{{\tt Goal 1.1.1} -- Sufficient Coverage of SPIs.}
There are two causes for potential insufficiencies of SPIs regarding coverage that need to be addressed: firstly, by its definition, an SPI is an an under-specified estimate of a claim's negation -- this means that if an SPI is false, then the claim is falsified. However, the claim can also be falsified when the SPI associated with it is true (the SPI "does not catch" all violations of the associated claim). Secondly, due to the large size of the safety case, only a subset of claims is feasible to be annotated with SPIs, and thereby, gaps can occur.


The sufficiency of the coverage of SPIs consists of three aspects. 
Firstly, enough lagging SPIs need to be defined to capture all violations of the item-level claims -- i.e., each incident, loss, and near miss at the system level is addressed by SPIs ({\tt Goal 1.1.1.1}). 
Secondly, the set of leading indicators needs to be enough to predict the losses in a timely manner such that the safety management system has time to react ({\tt Goal 1.1.1.2}). Thirdly, all important aspects of the safety case need to be sufficiently covered by SPIs ({\tt Goal 1.1.1.3}). The sub-goals of {\tt Goal 1.1.1.3} are mandated by clauses of UL4600 and can be extended with additional goals based on company standards or additional literature. We mark the fact that the set of sub-goals {\tt Goal 1.1.1.3.1 - 1.1.1.3.6} is incomplete by using three dots on the right of the argument leg.

\vspace{-0.2cm}
\paragraph{{\tt Goal 1.1.2} - Rigorous Management of SPIs.}
This argument leg addresses the "mechanics" of the management and implementation of SPIs. The chosen strategy is that the process to implement and manage SPIs is well defined and integrated with the other processes.
{\tt Goal 1.1.2.1} addresses the need for SPIs to be continuously monitored and updated to reflect safety case changes, system changes and environment changes. 
{\tt Goal 1.1.2.2} is about treating SPIs as an intrinsic part of configuration management. Again, there can be additional sibling claims of {\tt Goal 1.1.2.2},
illustrated with the three dots.

\subsection{Collection and Analysis of SPIs}
\label{sec:collection_and_analysis_of_spis}
\begin{figure}[h]
\vspace{-0.7cm}
\centering
\includegraphics[scale=0.55]{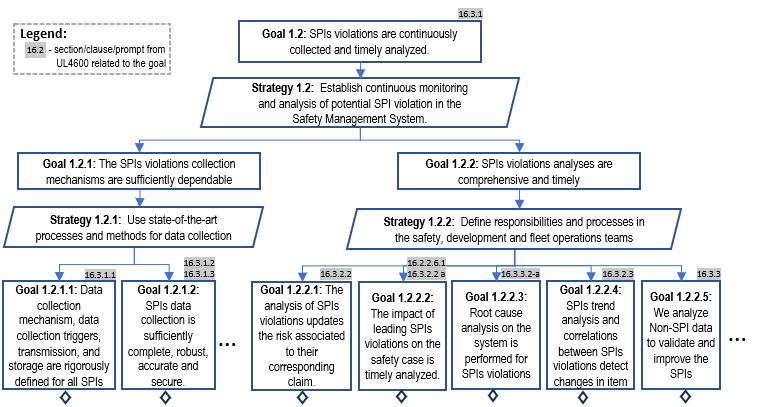}
\vspace{-0.3cm}
\caption{Argument leg about the collection and analysis of SPIs.}
\vspace{-0.8cm}
\label{fig:implicit_spis_safety_case_collection_and_analysis_of_spis}
\end{figure}

Once SPIs are defined and deployed to the fleet, their violations need to be collected and analyzed timely ({\tt Goal 1.2}) such that subsequent response actions can be enacted early enough to ensure the risk associated with the item remains acceptable. In \cref{fig:implicit_spis_safety_case_collection_and_analysis_of_spis} we present the argument leg supporting {\tt Goal 1.2}. This goal is developed in two sub-goals -- {\tt Goal 1.2.1} claiming the dependability of the collection mechanism for SPI violations, and {\tt Goal 1.2.2} claiming the comprehensiveness and timely analysis of violations of SPIs.

\paragraph{{\tt Goal 1.2.1} -- Dependable Collection of SPI Violations.} 
We need a dependable approach to collect SPI violations by clearly defining and robustly implementing the collection, triggering, data storage and transmission mechanisms ({\tt Goal 1.2.1.1} and {\tt Goal 1.2.1.2}) and potentially other sibling goals. 

\vspace{-0.2cm}
\paragraph{{\tt Goal 1.2.2} -- Analysis of SPI Violations }
Each SPI violation represents a violation of its associated safety claim and, thereby, a latent defect in the safety case, potentially leading to the falsification of the top claim, meaning unacceptable risk. Thereby, the analysis of violations shall be done in a comprehensive and timely manner. The timing aspect of the collection and analysis of violations is directly dependent on the exposure of the fleet. The strategy ({\tt Strategy 1.2.2}) to achieve this is to have clearly defined processes and responsibilities in the operations, safety, and product development teams. The violation of lagging SPIs needs to update the risk associated with the corresponding claim ({\tt Goal 1.2.2.1}). Violations of leading indicators ({\tt Goal 1.2.2.2}) need to be analyzed in a timely manner for the future impact and potential to produce losses. For each violation, a root cause analysis shall be performed ({\tt Goal 1.2.2.3}) in order to be able to define an adequate response.
Last but not least, trends in SPIs and correlations between non-SPI metrics shall be used by the safety engineers to detect changes in the item or environment and improve the set of SPIs ({\tt Goal 1.2.2.4, and 1.2.2.5}). 


\subsection{Response to SPI Violations}
\label{sec:response_to_spis_violations}

\begin{figure}[ht]
\centering
\includegraphics[scale=0.46]{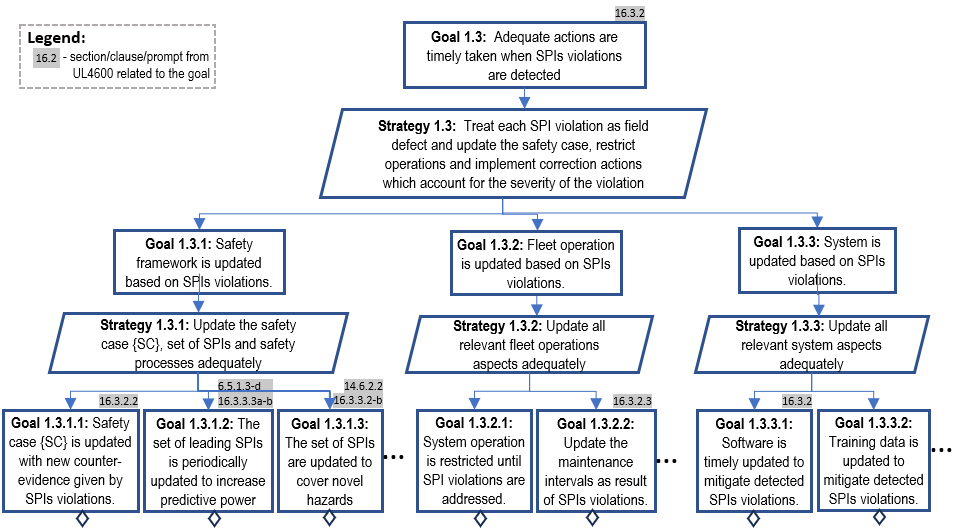}
\vspace{-0.5cm}
\caption{Argument leg about the response to SPIs violations.}
\label{fig:implicit_spis_safety_case_response_to_spis_violations}
\vspace{-0.5cm}
\end{figure}

After SPI violations are analyzed, an adequate response needs to be triggered -- {\tt Goal 1.3} from \cref{fig:implicit_spis_safety_case_response_to_spis_violations}. The response consists of updating the safety framework, updating fleet operations, and updating the system. Updating the safety framework, {\tt Goal 1.3.1} accounts for updating the safety case ({\tt Goal 1.3.1.1}), the set of leading SPIs to increase the prediction power ({\tt Goal 1.3.1.2}) or the lagging SPIs to cover newly identified hazards ({\tt Goal 1.3.1.3}).  

Updating the operations ({\tt Goal 1.3.2}) involves restricting the system operation ({\tt Goal 1.3.2.1}), updating maintenance intervals ({\tt Goal 1.3.2.2}), together with other potential actions.
The range of possible restrictions of the operation spans according to the determined risk increase due to the SPI violation -- it ranges from increasing safety margins by \eg reducing the vehicle speed, to restricting the operational domain by geofences or even grounding the fleet in an extreme case.
The time frame for implementing a response depends both on the fleet size and on the determined risk increase identified during the SPIs' violation analysis. 
%
The third response leg mandates an update of the system ({\tt Goal 1.3.3}), including software, ML training data, and potentially other aspects. 
\vspace{-0.2cm}

\section{Critical Analysis of the Argument about SPIs}\label{sec:critical_analysis}
Below, we present a set of potential challenges identified by performing a dialectic analysis of the proposed safety argument structure as introduced by \cite{2023_hawkins_identifying_runtime_monitoring_Requirements_for_autonomous_systems_through_analysis_of_safety_arguments}. 
Concretely, we look for how the claims of the argument presented in Section~\ref{sec:implicit_safety_case} could be undermined via \emph{challenge claims}~(CC) and in Section~\ref{sec:meta_spis} we search for what counter-evidence could undermine the truth of the argument. 
All claims can be challenged via challenge claims, but for brevity reasons, we focus only on a subset of the claims, which we consider to be subject of higher uncertainty. For example, we leave {\tt Goal 1.2.1.2} about the dependability of the SPI collection unchallenged since achieving it has a lower degree of uncertainty.

\vspace{-0.2cm}
\paragraph{Critical Analysis of the Argument Leg about the Definition of SPIs.}
Regarding the sufficiency of claims about the coverage of the definition of SPIs, we identified the following \enquote{challenge claims} (CC):

\emph{\emph{CC1)} Challenging {\tt Goal 1.1}}: Due to the high complexity of the system, the open world, and the dynamic nature of the environment, no set of SPIs can sufficiently detect validity violations of the safety case.

\emph{\emph{CC2)} Challenging {\tt Goal 1.1.1.1}}: Due to the under-approximative nature of SPIs regarding detection of claim violations, there can be unacceptably many violations of claims that are not covered by their corresponding SPIs.

\emph{\emph{CC3)} Challenging {\tt Goal 1.1.1.2}}: Due to dynamic environment and high system complexity, the prediction power of the chosen leading SPIs cannot be guaranteed a priori to be sufficient.

\emph{\emph{CC4)} Challenging {\tt Goal 1.1.1.3}}: Due to the big size of the safety case, not all argument legs of \{SC\} can be sufficiently covered with SPIs.


\vspace{-0.2cm}
\paragraph{Critical Analysis of the Argument Leg about the Collection and Analysis of SPIs Violations.}
Regarding the sufficiency of claims about the collection and analysis of SPI violations, we have identified the following challenge claims:

\emph{\emph{CC5)} Challenging {\tt Goal 1.2.2.2}}: Collection and analysis of leading SPI violations cannot be guaranteed to offer enough time to the safety, development, and operation teams to adequately respond before losses occur.

\emph{\emph{CC6)} Challenging {\tt Goal 1.2.2.3}}: Causes for certain SPI violations might not be identified timely and precisely enough to allow an adequate response.

\emph{\emph{CC7)} Challenging {\tt Goal 1.2.2.4}}: Due to the complexity of the system, no sufficient analysis of correlations between SPI violations can be guaranteed a priori to detect all relevant item changes.

\emph{\emph{CC8)} Challenging {\tt Goal 1.2.2.5}}:  Due to the high amount of data collected or potential gaps in the collected data, the non-SPI data cannot be guaranteed a priori to improve the SPIs sufficiently.



\vspace{-0.2cm}
\paragraph{Critical Analysis of the Argument Leg about the Response to SPIs Violations.}
Regarding the sufficiency of claims about the response to SPI violations, we have identified the following challenge claims.

\emph{\emph{CC9)} Challenging {\tt Goal 1.3}:} Once SPI violations occur, the response might be too late or ineffective to prevent losses. 


\emph{\emph{CC10)} Challenging {\tt Goal 1.3.1.2} and {\tt Goal 1.3.1.3}:} Due to high complexity and dynamics of the environment, the updated set of SPIs might not sufficiently increase the validation power for the safety case.

\emph{\emph{CC11)} Challenging {\tt Goal 1.3.2.1}:} System operation restriction, deployed as response to leading SPIs violations, might not be enough to avoid losses.



\section{Meta-SPIs to Validate the Argument about SPIs}\label{sec:meta_spis}

Once challenge claims are made explicit, we can pro-actively look for evidence that supports them -- i.e.,~check the validity of challenge claims as proposed by \cite{2023_hawkins_identifying_runtime_monitoring_Requirements_for_autonomous_systems_through_analysis_of_safety_arguments}. This evidence for challenge claims is at the same time counter-evidence for the argument for using SPIs. 
To tackle this, we propose to use special monitors (called meta-SPIs denoted \metaspi -- i.e., SPIs about the use of SPIs) to identify potential deficits. 
The set of meta-SPIs is incomplete and it represents only a starting point for maintaining the confidence in a certain set of SPIs.

\vspace{-0.2cm}
\paragraph{Monitors for the Validity of the Argument Leg about Coverage of SPIs.} Below we define a set of \metaspi s which detect deficits in the argument about the sufficiency of SPIs coverage (the leg corresponding to {\tt Goal 1.1}).

\emph{\metaspi1 for CC1:}
This monitor detects cases when system level losses are occurring, but no SPI augmenting the lower level parts of the safety case \{SC\} is violated. This means that despite instrumenting the safety case with SPIs, they rendered to be insufficient in providing early warnings and preventing loses. 

\emph{\metaspi2 for CC2, CC3, CC4:}
This monitor detects cases when SPIs associated with higher-level claims of \{{\tt SC}\} are falsified, but no SPIs associated to lower-level claims have been falsified. In this case, the prediction power of lower-level SPIs is insufficient, and the definition of additional lower-level SPIs is required.


\vspace{-0.2cm}
\paragraph{Monitors for the Validity of the 'Collection and Analysis' Argument Leg.} Below we define a set of \metaspi s which detect deficits in the argument about the sufficiency of collection and analysis of SPIs violations (the leg of  {\tt Goal 1.2}).

\emph{\metaspi3 for CC5:}
We define a monitor that detects cases when violations of the same SPIs associated to lower level claims are reported before the completion of analysis of previous violations of these SPIs. In this case, the SMS is trying to catch up and is saturated by the set of reported violations. Thereby, immediate contingency actions need to be deployed.

\emph{\metaspi4 for CC6:}
We define a monitor that detects cases when no conclusive causes could be found when a certain SPI was violated, and thereby, no sufficient response is possible to implement. This requires increasing of diagnostics capabilities of the system.

\emph{\metaspi5 for CC7 and CC8:} 
We define a monitor that detects cases when violations of SPIs attached to lower level claims keep occurring but the correlations of these violations or additionally collected data do not offer insights about changes of the system or possible weaknesses of the chosen set of SPIs.

\vspace{-0.2cm}
\paragraph{Monitors for the Validity of the 'Response to SPIs Violations' Argument Leg.} Below we define a set of \metaspi s which detect deficits in the argument about the sufficiency of response to SPIs violations (the leg corresponding to {\tt Goal 1.3}).

\emph{\metaspi6 for CC9:} 
We define a monitor that detects cases when deploying a response for a certain SPI violation does not sufficiently prevent further violations of the same SPI. The implemented response rendered to be ineffective. 

\emph{\metaspi7 for CC10:} 
We define a monitor that detects cases when the updated set of SPIs associated to lower level claims does not improve the prediction power and SPIs violations associated to higher level claims keep being a surprise. This situation puts the prediction power of these SPIs into question and signalizes future surprises in the corresponding argument leg.

\emph{\metaspi8 for CC11:} 
We define a monitor that detects cases when the restricted system operations following the violation of an SPI does not prevent further violations of that SPI or SPIs attached to higher-level claims in the safety case. In this case, tighter restrictions need to be immediately deployed following even a potential grounding of the fleet.



\section{Related Work}\label{sec:related_work}


\paragraph{Dynamic Safety Cases and Perpetual Assurance of Adaptive Systems.} 
Foundational research work has been performed on the assurance of adaptive systems.
Denney, Pai, and Habli~\cite{2015_denney_dynamic_safety_cases_for_through_life_safety_assurance} coined the notion of dynamic safety cases as an engineering solution to achieve through-life safety assurance and enable proactive safety management. 
Weyns et al.~\cite{2017_weyns_perpetual_assurance_for_self_adaptive_systems} defined a continuous process where evidence is gathered and integrated into the safety case and, whenever needed, leads to updates of the safety case. 

However, the foundational research does not describe how to concretely define the monitors, collect the needed information from runtime or how to link violations to the safety case or how to respond. 

\paragraph{Monitoring of the Safety Case of Autonomous Systems.} 
In the context of autonomous systems, there is a broad acceptance of the need to monitor the safety case to bridge the gap between design time assumptions and actual performance.
McDermid et al.~\cite{2019_mcdermid_framework_for_safety_assurance_of_autonomous_systems} emphasizes the need to continuously update the safety case (initially based on assumptions about the \enquote{world as imagined}) with information from operations (rendering the \enquote{world as observed}). 
Gyllenhammar, Bergenhem and Warg~\cite{2021_gyllenhammar_ads_safety_assurance_future_directions} emphasize the need to use data from operations to ensure the validity of the safety case of autonomous systems;
Johanssonn and Koopman~\cite{2022_johannsson_continuous_learning_approach_to_safety_engineering} propose that real-world field feedback be incorporated into safety cases.
The standard UL~4600\cite{2023_ul4600} mandates the use of Safety Performance Indicators for gathering potential violations of the safety claims in a systematic manner and taking adequate actions to mitigate the increased risks. 
The UL-4600 guidelines are used as basis for the development of safety case frameworks \cite{2024_wagner_open_autonomy_safety_case_framework} which  use SPIs as a central pillar to check the validity of the safety case. 
        
To the best of our knowledge, there is no published safety case pattern specifically focused on the use of runtime monitors associated with safety cases in general and the use of SPIs as defined by UL~4600 in particular. While the definition of runtime monitors has been closer investigated in the AV safety research community, there is no explicit argument about their use for maintaining the validity of the safety case. Furthermore, there is no systematic discussion about the challenges of deploying SPIs, investigation of their potential deficits and how these deficits could be addressed. 

\section{Summary and Future Work}\label{sec:conclusions}
The validity of safety cases of autonomous systems needs to be maintained throughout their entire lifecycle. SPIs are mandated by UL~4600~\cite{2023_ul4600} to be used to monitor the validity of the safety cases, and in case when violations are detected to analyze their impact and perform adequate response. In this paper, we present an argument structure about the use of SPIs to maintain the validity of a safety case. We make explicit the rationale and implicit argument about the role of SPIs in order to increase the awareness of all involved stakeholders. We analyze the claims of our argument structure for potential defeaters and identify a set of challenge claims that would invalidate the SPI approach. We propose a set of monitors, called meta-SPIs, which can be used to validate the SPIs framework.

This work represents only an initial step to systematically define an SPI framework to ensure the continuous validity of safety cases. As future work we plan to collect feedback about our argument from all relevant stakeholders, extend the coverage of the argument to address their needs and evolve it into a pattern. Furthermore, we plan to validate the use meta-SPIs as means to provide confidence in the SPIs framework.
Our longer term goal is to automate this process and support it with tooling in order to be able to deploy it at scale. 

\vspace{-0.3cm}
\section*{Acknowledgment}
\vspace{-0.1cm}
This work was partially supported by the German Federal Ministry of Education and Research in the project \textit{MANNHEIM-AutoDevSafeOps} (011S22087R).

\vspace{-0.3cm}

\bibliographystyle{splncs04}
\bibliography{bibliography}

\end{document}